\documentclass[12pt]{iopart}

\usepackage{graphicx}  

\begin{document}

\paper[Analytical calculation of neighborhood order probabilities \ldots]{Analytical calculation of neighborhood order probabilities for high dimensional Poissonic processes and mean field models}

\author{C\'{e}sar  Augusto Sangaletti Ter\c{c}ariol$^1$$^{\mbox{,}}$$^2$, Felipe de Mouta Kiipper$^2$ and Alexandre Souto Martinez$^2$}

\address{$^1$ Centro Universit\'ario Bar\~ao de Mau\'a \\
             Rua Ramos de Azevedo, 423 \\ 
             14090-180, Ribeir\~ao Preto, SP, Brazil}

\address{$^2$ 
  Faculdade de Filosofia, Ci\^encias e Letras de Ribeir\~ao Preto, \\
  Universidade de S\~ao Paulo \\ 
  Avenida Bandeirantes, 3900 \\ 
  14040-901, Ribeir\~ao Preto, SP, Brazil.
}
\eads{\mailto{cesartercariol@gmail.com}, \mailto{fekiipper@yahoo.com}, \mailto{asmartinez@ffclrp.usp.br}}

\begin{abstract}
Consider that the coordinates of $N$ points are randomly generated along the edges of a $d$-dimensional hypercube (random point problem).
The probability that an arbitrary point is the $m$th  nearest neighbor to its own $n$th nearest neighbor (Cox probabilities) plays an important role in spatial statistics.
Also, it has been useful in the description of physical processes in disordered media.
Here we propose a simpler derivation of Cox probabilities, where we stress the role played by the system dimensionality $d$. 
In the limit $d \rightarrow \infty$, the distances between pair of points become indenpendent (random link model) and closed analytical forms for the neighborhood probabilities are obtained both for the thermodynamic limit and finite-size system.
Breaking the distance symmetry constraint drives us to the random map model, for which the Cox probabilities are obtained for two cases: whether a point is its own nearest neighbor or not.
\end{abstract}


\submitto{\JPA}
\maketitle

\section{Introduction}

Consider $N$ points independent and uniformly distributed along the edges of a $d$-dimensional hypercube.
The determination of the distance and neighborhood statistics between any pair of points is known as the \emph{random point problem} (RPP).
This is a standard approach to construct disordered (random) media.

Due to boundary effects and triangular restrictions, the distances between any pair of points are not all independent random variables.
For fixed $N$ in the RPP, as the system dimensionality $d$ increases, the boundary effects become more and more pronounced and the distances between pair of points become less and less correlated.
One can minimize boundary effects considering periodic boundary condition, and in the limit $d \rightarrow \infty$ all the two--point distances are independent and identically distributed (i.i.d.) random variables.
This is the \emph{random link (distance) model} (RLM)~\cite{mezard:1986}, which is a mean field description of the RPP.

In the RLM, there exist two Euclidean constraints: (i) the distance from a point to itself is always null ($D_{ii} = 0$, for all $i$) and (ii) the forward and backward distances are equal ($D_{ij} = D_{ji}$, for all $i$, $j$).
The second constraint imposes serious numerical difficulties and an efficient numerical implementation for the RLM is given in Ref.~\cite{tercariol_2006}.
If the distance symmetry constraint is broken, the model becomes the \emph{random map model} (RMM)~\cite{harris:1960,derrida:2:1997}.
In this latter model, a point can be whether its own nearest neighbor ($D_{ii} = 0$) or not ($D_{ii} \neq 0$).
The latter is the mean field approximation for Kauffman automata~\cite{kauffman:1969}.

Both, the RPP and RLM have been very fruitful in the determination of numerical and analytical results in several interesting systems.
Applications range from statistics on the optimal trajectories in the context of traveling salesman problem on a random set of cities~\cite{percus:1996,percus:1997,percus:1999,aldous:2003,aldous:2005}, passing by frustrated dimerization optimization modeled by the minimum matching problem~\cite{Boutet:1997,Boutet:1998} (or equivalently spin-glasses~\cite{Boutet:1997}), and going to partial self-avoiding deterministic tourist walk~\cite{lima_prl2001,stanley_2001,kinouchi:1:2002,tercariol_2005} and its random version~\cite{risaugusman:1:2003,martinez:1:2004}.
Partial self-avoiding walks have been our main motivation to address the RPP and its mean field models.
Although the distance distribution as a function of the dimensionality $d$ plays an important role in the random tourist version, in the deterministic case one is mainly interested on the neighborhood ranking of random points.

As pointed above, boundary effects are important as the dimensionality of the system increases. 
The points get closer to the surface and to capture the bulk effect, one must increase $N$. 
In certain systems it may be difficult to have such large $N$ values and it would be suitable to have analytical expressions for finite $N$, for instance, to test reliability of numerical codes or to develop new statistical tests.

Here we focus on the distribution of neighborhood ranks. 
The probability that an arbitrary point is the $m$th nearest neighbor of its own $n$th nearest neighbor in the RPP has attracted attention of researchers since the seminal studies of Clark and Evans~\cite{clark_1955} and Clark~\cite{clark_1956}.
They devised the term reflexive neighbors for the case $m=n$ and their calculated reflexive neighborhood probability ranking has been corrected by Dacey ($m > 1$) and then generalized (for $m \ne n$) by Cox~\cite{cox}, which we call the \emph{Cox probabilities}.

In this paper, in Sec.~\ref{sec:cox_probs} we obtain the Cox probabilities using only Poisson distribution instead of the various distinct distributions used in the original paper~\cite{cox}.
As in Cox calculation, we write the probabilities in the thermodynamic limit $N \rightarrow \infty$. 
Unlike Cox, we write them in terms of known functions (rather than in terms of an integral). 
In Sec.~\ref{sec:rlm}, the use of known special functions allows us to take the high dimensionality limit,  which leads to the RLM neighborhood probability. 
Using the same arguments to obtain Cox probabilities, we are abke to obtain neighborhood probability for finite-size RLM systems. 
Finally, in Sec~\ref{sec:rmm} we explicitly write the Cox probabilities for the two considered case of the RMM.
After the concluding remarks~(\ref{sec:conc}), the discrete probability distributions used are briefly reviewed to set up notation in the Appendix. 
All analytical results have been compared and validated by numerical Monte Carlo simulations.

\section{Alternative Derivation of Cox Probabilities}
\label{sec:cox_probs}

This alternative derivation of Cox formula is simpler than the original paper, since it uses only the Poisson distribution, rather then the Poisson, binomial and gamma distributions as in the original paper.

In a $d$-dimensional Poissonic medium with an mean density of $\lambda_d$ points per unitary volume, the probability that a volume $V_d$ (with an arbitrary shape, even with disconnected parts) contains $k$ points is given by the Poisson distribution $\mbox{pois}(k)  =  \mu^k e^{-\mu}/k!$, where $k = 0, 1, 2, \ldots, \infty$ and $\mu = \langle k \rangle = \lambda_d V_d$ is the expected number of points inside the volume $V_d$.
Notice that the thermodynamic limit is taken letting $k$ freely vary and that the only parameter of this distribution is $\mu$ (the medium dimensionality $d$ is not a relevant quantity).

\begin{figure}[htb]
\begin{center}
\includegraphics[angle=-90, scale=.5]{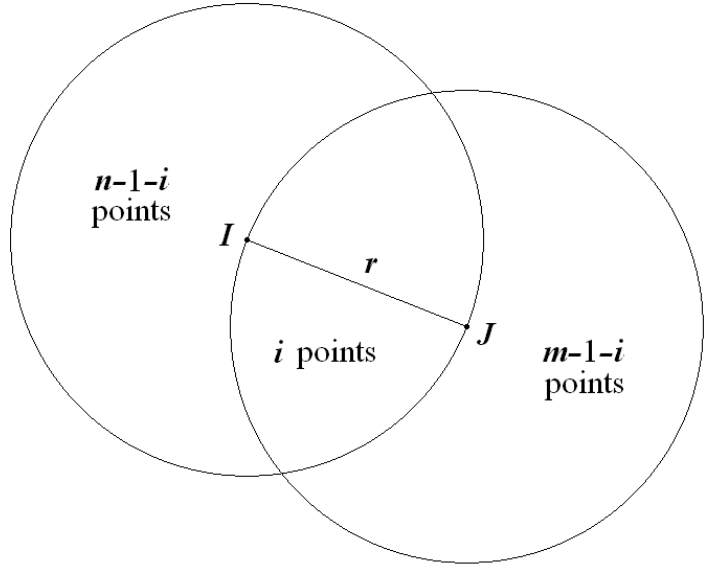}
\caption{Two-dimensional poissonic process. The circles centered in the points $I$ and $J$ have surface $V_2 = \pi r^2$ and the intersection has an area $V_{\cap,2} = V_2(1-p_2) = (2\pi/3 - \sqrt{3}/2) r^2$. There are $i$ points in the intersection of the $V_2$ surfaces and in the $I$ and $J$ crescents there are $n-1-i$ and $m - 1 -i $ points, respectively.} 
\label{Fig:Hiperesferas}
\end{center}
\end{figure}

Let $I$ and $J$ be two points of a $d$-dimensional Poissonic medium separated by a distance $r$.
The volume $V_d(r)$ of the hypersphere of radius $r$ centered in $I$ (thus, which pass through $J$) is $
V_d(r)  =  \pi^{d/2} r^d/\Gamma(d/2+1) = \pi^{(d-1)/2} r^d  B[1/2, (d+1)/2]/\Gamma[(d+1)/2]$, 
where $\Gamma(z) = \int_0^{\infty} \mbox{d}t \; t^{z-1}e^{-t}  = (z-1)!$ is the \emph{gamma function}~\cite{stegun} and $B(a,b) = B(b,a) = \int_{0}^{1} \mbox{d}t \; t^{a-1}(1-t)^{b-1} = \Gamma(a) \Gamma(b)/\Gamma(a+b)$ is the \emph{beta function}~\cite{stegun}.
While the former generalizes the factorial the latter is a generalization of the inverse of Newton binomial.
Obviously the volume of the hypersphere centered in $J$ passing through $I$ is also $V_d(r)$.
Fig.~\ref{Fig:Hiperesferas} shows the case $d=2$.

The volume $V_{\cap,d}(r)$ of the intersection of these two hyperspheres is $
V_{\cap,d}(r)  =  \{\pi^{(d-1)/2} r^d/\Gamma[(d+1)/2]\} \int_{1/4}^1 \mbox{d}t \; t^{-1/2} (1-t)^{(d-1)/2}$.  
The relative volume of a crescent (compared to one hypersphere) is $p_d  =  [V_d(r)-V_{\cap,d}(r)]/V_d(r) = \int_0^{1/4} \mbox{d}t \; t^{-1/2} (1 - t)^{(d-1)/2}/B[1/2, (d+1)/2]$ or: 
\begin{equation}
p_d  =  I_{1/4}\left(\frac{1}{2}, \frac{d+1}{2} \right)
\label{Eq:pd1}
\end{equation}
where $I_z(a,b) =  \int_{0}^{z} \mbox{d}t \; t^{a-1}(1-t)^{b-1}/B(a,b)$ with $\mbox{Re}(a) > 0$, $\mbox{Re}(b) > 0$ is the {\em normalized incomplete beta function}~\cite{stegun}.
Notice that $p_d$ depends exclusively on the dimensionality $d$ and does not depend on the hypersphere radius $r$.

It is interesting to mention that $p_d$ plays an important role in the parametrization of the deterministic tourist walk problem~\cite{kinouchi:1:2002,tercariol_2005}. 
It can be generalized to an arbitrary distance $D_{I,J} = rx$ between the points $I$ and $J$, with $x$ ranging from 0 to 2 (from concentric hyperspheres to disjoined ones).
In this case, one has $p_d(x) = I_{(x/2)^2}[1/2,(d+1)/2]$, with $p_d(1) = p_d$, which has allowed us to tackle analytically the random tourist walk problems~\cite{risaugusman:1:2003,martinez:1:2004}.

The following conditions must hold for $I$ be the $m$th nearest neighbor of $J$ and $J$ be the $n$th nearest neighbor of $I$:
\begin{enumerate}
\item there must exist $i$ points inside the intersection of the hyperspheres, with $i$ ranging from 0 to $\mbox{min}(m-1, n-1)$, the expected number of points is $\mu (1-p_d)$,
\item there must exist $m-1-i$ points inside the crescent of $J$, the expected number of points is $\mu p_d$,
\item there must exist $n-1-i$ points inside the crescent of $I$, the expected number of points is also $\mu p_d$,
\item the distance $r$ between $I$ and $J$ may assume any value in the interval $[0,\infty)$, allowing the volume $A_d(r)$ and expected number of points $\mu = \lambda A$ inside it also vary from 0 to $\infty$ (continuous value).
\end{enumerate}

Taking these conditions altogether, one obtains the following expression for the probability $P^{(d)}_{m,n}= P^{(d)}_{n,m}$:
\begin{eqnarray*}
P^{(d)}_{m,n}
& = & \int_0^\infty \mbox{d}\mu \sum_{i=0}^{\mbox{\footnotesize min}(m-1, n-1)} \frac{[\mu(1-p_d)]^i e^{-\mu(1-p_d)}}{i!}\cdot \\ 
&   & \frac{(\mu p_d)^{m-1-i} e^{-\mu p_d}}{(m-1-i)!} \cdot \frac{(\mu p_d)^{n-1-i}e^{-\mu p_d}}{(n-1-i)!} \; .
\end{eqnarray*}
Collecting the factors which do not depend on $\mu$, the integral can be written in terms of the gamma function: 
$\int_0^\infty \mbox{d}\mu \; \mu^{m+n-2-i} e^{-\mu(1+p)}  =  \Gamma(m+n-1-i)/(1+p)^{m+n-1-i}$
 and one obtains the original form of Cox probabilities: 
\begin{eqnarray}
P^{(d)}_{m,n} & = & \sum_{i=0}^{\mbox{\footnotesize min}(m-1, n-1)} \frac{(m+n-2-i)!}{i!(m-1-i)!(n-1-i)!} 
\nonumber \\ & & 
\frac{(1-p_d)^i p_d^{m+n-2-2i}}{(1+p_d)^{m+n-1-i}}
\label{Eq:FormulaCox1}
\end{eqnarray}
with $m = 1, 2, \ldots, \infty$ and $n = 1, 2, \ldots, \infty$.
Letting $i$ vary from 1 to $\mbox{min}(m,n)$ and rearranging the terms, the summed expression may be identified with the multinomial distribution.
\begin{eqnarray}
\frac{P^{(d)}_{m,n}}{P^{(d)}_{1,1}} & = & \sum_{i=1}^{\mbox{\footnotesize min}(m, n)} \mbox{mult}\left(i-1, m-i, n-i; \frac{1-p_d}{1+p_d},  \frac{p_d}{1+p_d}, \frac{p_d}{1+p_d}\right)
\label{Eq:CoxAltern} \\ 
P^{(d)}_{1,1} & = & \frac{1}{1+p_d} \; ,
\label{Eq:CoupleDensity1}
\end{eqnarray}
where $P^{(d)}_{1,1}$ is the couple density (mutually nearest neighbors) and 
$\mbox{mult}(n_a, n_b, n_c; p_a, p_b, p_c)$ is the multinomial distribution (Eq.~\ref{Eq:MultinomialDistribution}).  

Notice that the Cox probability distribution is not a joint distribution.
The summation $\sum_{m,n} P^{(d)}_{m,n}$ diverges since for each neighborhood degree $m$ it must be normalized $\sum_{n} P^{(d)}_{m,n} = 1$ and one obtains the mean $\langle n \rangle =m + p_d$ and the variance $\langle n^2 \rangle - \langle n \rangle^2 =(2m + p_d - 1)p_d$. 
The system dimensionality $d$ is the bare parameter that emerges from the medium while the considered neighborhood order $m$ is fixed according to the convenience.   

For future reference, let us rewrite Eq.~\ref{Eq:FormulaCox1} close to its original form:
\begin{eqnarray}
\nonumber
P^{(d)}_{m,n} & = & \frac{1}{1-p_d} \left( \frac{p_d}{1 + p_d} \right)^{m+n} 
\\ & & \nonumber
\sum_{i=0}^{\min\{m-1,n-1\}} \frac{[(1 - p_d^2)/p_d^2]^{i+1}(m + n - i - 2)! }{(m - i - 1)! (n - i - 1)! i! } \\ 
& = & \frac{(p_d^{-1} + 1)^{-(m+n)}}{1-p_d} 
\nonumber \\ & & 
\sum_{i=1}^{\min\{m,n\}} 
 \frac{( p_d^{-2} - 1 )^{i} B(m+n-2i,2) }{B(m+n-2i,i) B(m-i+1,n-i+1)}  \; .
\label{Eq:FormulaCox3}
\end{eqnarray}

Numerical values of Eqs.~\ref{Eq:pd1} and~\ref{Eq:FormulaCox1} are shown in Table~\ref{Tab:Geral2}.

\section{Random Link Model and High Dimensionality Probabilities}
\label{sec:rlm}

The high dimensionality can be obtained directly from Cox probabilities. 
It is interesting to point out the emergence of a characteristic dimensionality $d_0 = 8$ from the calculation. 
In this procedure, one can easily obtain the first order correction from the random link model neighborhood probabilities. 
Next we recall that we are considering the thermodynamic limit and give a geometrical interpretation for the random link expression, which corresponds to all the points being on the surface of the volume $V_d$. 
In the following we correct the random link model neighborhood probabilities to finite systems.

\subsection{Thermodynamic Limit}

Let us consider the high dimensionality situation ($d \gg 1$).
This corresponds to take $b=(d+1)/2 \gg a=1/2$ in Eq.~\ref{Eq:pd1}.
Since $b \gg a$, the approximation $ B(a,b) \approx \Gamma(a)/b^a$ can be used for $I_z(a,b) \approx b^a/\Gamma(a) \int_0^{z} \mbox{d}t \;  t^{a-1} (1 - t)^b$.
Once $t \le z = 1/4$ implies $t \ll 1$, the approximation $(1 - t)^b = e^{b \ln(1 - t)} \approx e^{-bt}$ yields
$I_z(a,b) \approx \gamma(a, bz)/\Gamma(a)$, where $\gamma(a,b) = \int_0^{b} \mbox{d}t \; t^{a-1}e^{-t}$ is the \emph{non-normalized incomplete gama function}~\cite{stegun}, which presents the following property
$\gamma(1/2,x) = 2 \int_0^{\sqrt{x}} \mbox{d}t \; e^{-t^2} = \sqrt{\pi} \mbox{erf}(\sqrt{x})$
with the \emph{error function}~\cite{stegun} definided by:
$\mbox{erf}(z) = (2/\sqrt{\pi}) \int_0^z \mbox{d}t \; e^{-t^2}$ 
which monotoly increases from $\mbox{erf}(0) = 0$ to $\mbox{erf}(\infty) = 1$.
Since $a = 1/2$, the following property~\cite{stegun} can be used: $I_z(a,b) \approx \gamma(1/2, bz)/\Gamma(1/2) = \mbox{erf}(\sqrt{bz})$
and  Eq.~\ref{Eq:pd1} can be re-written as: 
\begin{equation}
p_d  \approx  \mbox{erf}\left(\sqrt{\frac{d}{8}}\right) \; .
\end{equation}
where a characteristic dimensionality $d_0 = 8$ naturally emerges from the analysis.

The {\em complementar error function} is defined by
$\mbox{erfc}(z) = (2/\sqrt{\pi}) \int_z^\infty \mbox{d}t \; e^{-t^2} = 1 - \mbox{erf}(z)$.
For $|z| \gg 1$, its Taylor series is useful: $\mbox{erfc}(z) = e^{-z^2}/(z \sqrt{\pi}) ( 1 - z^2/2 + \cdots )$.  
A further approximation can be performed noticing that $\mbox{erfc}(z) = 1 - \mbox{erf}(z)$, so that for $|z| \gg 1$, it  can be written as~\cite{stegun}:
\begin{equation}
p_d \approx  1 - \mbox{erfc}\left(\sqrt{\frac{d}{8}}\right) \approx  1 - \alpha_d \; ,
\label{Eq:PdAssint}
\end{equation}
with
\begin{equation}
\label{Eq:AlfaD}
\alpha_d  =  \frac{1}{\sqrt{\pi}} \; \frac{e^{-d/8}}{\sqrt{d/8}} \; \left(1 - \frac{4}{d}  + \cdots \right) \; .
\end{equation}

Using: $1-p_d  =  \alpha_d$, $p_d^{-1} + 1  =  2$ and $p_d^{-2} - 1  =  2 \alpha_d $ in Eq.~\ref{Eq:FormulaCox3}, Cox probabilities are written as a power series in $\alpha_d$ for high dimensional systems: 
\begin{equation}
P^{(d \gg 1)}_{m,n} = P^{(rl)}_{m,n} + \frac{2^{2-(m+n)}}{B(m-1,n-1)} \; \alpha_d + \cdots \; , 
\end{equation}
where in the random link approximation $(d \rightarrow \infty)$ this probability is:
\begin{eqnarray}
\nonumber
\frac{P^{(rl)}_{m,n}}{P^{(rl)}_{1,1}} & = & \frac{1}{2^{m+n}} \; \frac{\Gamma(m+n-1)}{\Gamma(m)\Gamma(n)}  = \frac{2^{-(m+n)}}{(m + n - 1) B(m,n)} \\
               & = &  \mbox{bin}\left(m-1, n-1, \frac{1}{2}, \frac{1}{2}\right)
\label{Eq:CoxRandomLink} \\
P^{(rl)}_{1,1} & = & \frac{1}{2} \; , 
\label{Eq:CoupleDensityRL}
\end{eqnarray}
where $P^{(rl)}_{1,1}$ is the couple density and $\mbox{bin}(n_a, n_b; p_a, p_b)$ is the binomial distribution given by Eq.~\ref{Eq:BinomialDistribution}.
Simple expressions can be obtained such as: $P^{(rl)}_{1,n}  = 1/2^{n}$, $P^{(rl)}_{2,n}  = n/2^{n+1}$.

In the high dimensionality limit $d \rightarrow \infty$, the relative volume of the crescent (Eq.~\ref{Eq:pd1}) tends to 1 ($p_d \rightarrow 1$) and the expected number of points $\mu(1-p_d)$ inside the intersection vanishes, 
This is easily seen if one considers a hypersphere of radius $r$ inside in a hypercube of edge $2r$, as the dimensionality increases the hypersphere volume decreases relatively to the hypercube and difference of volumes increases meaning that all the points lie on the external volume to the hypersphere~\cite{rodrigo_msc}.

Since $\lim_{p_d \rightarrow 1} [(1-p_d)/(1+p_d)]^i = \delta_{i,0}$, where $\delta_{i,j}$ is the Kronecker delta, the multinomial distribution in Eq.~\ref{Eq:CoxAltern} becomes the binomial distribution of Eq.~\ref{Eq:CoxRandomLink}.

The numerical values relative to the high dimensionality cases are shown in Table~\ref{Tab:Geral2}. 

\begin{table*}[ht]
\begin{center}
\begin{tabular}{c|c|ccc}
\hline
\hline
$d$      & $p_d$ & $P^{(d)}_{1,1}$         & $P^{(d)}_{1,2}$                                       & $P^{(d)}_{2,2}$ \\ 
\hline
\hline
$0$      &1/3 &$3/4$                   & $3/16$                                        & $15/32$ \\
\hline
$1$      &1/2 & $2/3$                  & $2/9$        & $10/27$  \\
\hline
$2$      &$\frac{2\pi + 3\sqrt{3}}{6 \pi}$ &$\frac{6\pi}{8\pi + 3\sqrt{3}}$ & $\frac{6\pi(2\pi + 3\sqrt{3})}{(8\pi + 3\sqrt{3})^2}$ & $\frac{6\pi(40 \pi^2 + 12 \sqrt{3} \pi + 27)}{(8\pi + 3\sqrt{3})^3}$ \\
\hline
$3$      &11/16 &$16/27$                 & $176/729$                                     & $6032/19683$ \\
\hline
$\vdots$ & &$\vdots$                & $\vdots$ & $\vdots$ \\
\hline
$\gg 1$  & $1 - \alpha_d$           & $(1+p_d)^{-1}$ & $p_d(1 + p_d)^{-2}$ & $(1+p_d^2)(1 + p_d)^{-3}$  \\
\hline
$\infty$ (rl) & 1 & $1/2$                   & $1/4$    & $1/4$ \\
\hline
\hline
\end{tabular}
\end{center}
\caption{Some values of neighborhood probability.
For low dimensionalities, one uses Eq.~\ref{Eq:FormulaCox1}.
An interesting limiting case is $d = 0$, which yields: $p_0 = \int_0^{1/4} \mbox{d}t/[\pi \sqrt{t(1-t)}] = 1/3$.
For $d \gg 1$, one uses Eq.~\ref{Eq:PdAssint} and for the random link model $d \rightarrow \infty$, one uses Eq.~\ref{Eq:CoxRandomLink}.}
\label{Tab:Geral2}
\end{table*}

\subsection{Finite Size System}

The RPP high dimensional limit $d \rightarrow \infty$ corresponds to the RLM, where all distances become i.i.d. random variables.
Since Euclidean distances are only a means to obtain the ranking neighborhood probabilities, it is independent of particular choice for the distance probability distribution function (pdf)~\cite{tercariol_2005}.
For simplicity, we will consider uniform deviates in the interval $[0, 1]$ for the distances among the $N$ points.

\begin{figure}[htb]
\begin{center}
\setlength{\unitlength}{.5mm}
\begin{picture}(136, 90)

\put(40, 43){\circle*{2}}
\put(35, 41){$I$}
\put(90, 43){\circle*{2}}
\put(93, 41){$J$}
\put(40, 43){\line(1, 0){50}}
\put(65, 44){$x$}

\put(40, 43){\line(1, 6){4}}
\put(44, 67){\circle*{1}}

\put(40, 43){\line(-1, 4){5}}
\put(35, 63){\circle*{1}}

\put(40, 43){\line(2, 5){10}}
\put(50, 68){\circle*{1}}

\put(30, 58){\dashbox{.5}(25, 15)}
\put(11, 85){$N-n-1$ points}
\put(15, 77){farther than $J$}

\put(40, 43){\line(-1, -6){4}}
\put(36, 19){\circle*{1}}

\put(40, 43){\line(1, -6){3}}
\put(43, 25){\circle*{1}}

\put(40, 43){\line(1, -2){10}}
\put(50, 23){\circle*{1}}

\put(30, 15){\dashbox{.5}(25, 15)}
\put(18, 8){$n-1$ points}
\put(17, 0){nearer than $J$}

\put(90, 43){\line(1, 6){4}}
\put(94, 67){\circle*{1}}

\put(90, 43){\line(-1, 4){5}}
\put(85, 63){\circle*{1}}

\put(90, 43){\line(-2, 5){10}}
\put(80, 68){\circle*{1}}

\put(74, 58){\dashbox{.5}(25, 15)}

\put(75, 85){$N-m-1$ points}
\put(75, 77){farther than $I$}

\put(90, 43){\line(-1, -6){3}}
\put(87, 58){\circle*{1}}

\put(90, 43){\line(1, -6){4}}
\put(94, 19){\circle*{1}}

\put(90, 43){\line(1, -2){10}}
\put(100, 23){\circle*{1}}

\put(81, 15){\dashbox{.5}(25, 15)}

\put(80, 8){$m-1$ points}
\put(80, 0){nearer than $I$}

\end{picture}
\caption{Schematic illustration of the points $I$ and $J$ and their neighbors in a $N$-point random link model.}
\label{Fig:CoxRandomLink}
\end{center}
\end{figure}
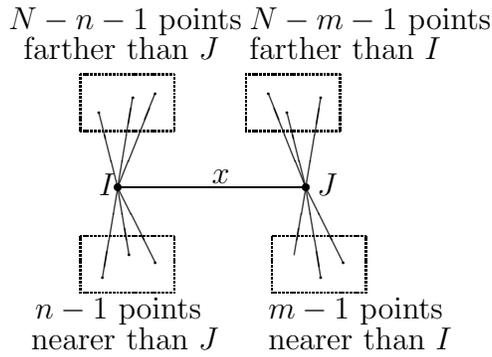

As before, let $I$ be the $m$th nearest neighbor of $J$ and $J$ be the $n$th nearest neighbor of $I$.
Thus, the following conditions hold:
\begin{enumerate}
\item the distance $x$ from $I$ to $J$ may assume any value in the interval $[0,1]$,
\item the distances from $J$ to each of its $m-1$ nearest neighbors must be less than $x$ and
\item the distances from $J$ to each of its $N-m-1$ farthest neighbors must be greater than $x$, as well as
\item the distances from $I$ to each of its $n-1$ nearest neighbors must be less than $x$ and
\item the distances from $I$ to each of its $N-n-1$ farthest neighbors must be greater than $x$.
\end{enumerate}
Figure~\ref{Fig:CoxRandomLink} illustrates the situation.

It also must be noticed that:
\begin{enumerate}
\item choosing an arbitrary point $I$, its $m$th nearest neighbor is automatically set, and there is $N-1$ possibilities for this,
\item it must be counted all possible combinations in distributing the $N-2$ neighbors of $J$ between the $m-1$ nearest and the $N-m-1$ farthest than $J$,
\item the same counting must be done for the $N-2$ neighbors of $I$.
\end{enumerate}

Combining these three countings and those five distance restrictions, one has:
\begin{eqnarray*}
P^{(rl, N)}_{m, n}
& = & \frac{(N-1)[(N-2)!]^2}{(m-1)!(N-m-1)!(n-1)!(N-n-1)!} \cdot 
 \\  &   & 
\int_0^1 \mbox{d}x \left[\int_0^x \mbox{d}y \right]^{m+n-2} \cdot \left[\int_x^1 \mbox{d}y \right]^{2N-m-n-2}
\end{eqnarray*}
Since: $\int_0^1 \mbox{d}x \; x^{m+n-2}(1-x)^{2N-m-n-2} = B(m+n-1, 2N-m-n-1) = (m+n-2)!(2N-m-n-2)!/[(2N-3) (2N-4)!]$ then:
\begin{eqnarray}
\frac{P^{(rl, N)}_{m, n}}{P^{(rl, N)}_{1, 1}} & = & 
\mbox{hypg}(N-2, N-2; m-1, n-1)  
\label{Eq:CoxRL_N} \\
P^{(rl, N)}_{1, 1} & = & \frac{N-1}{2N-3} \; ,
\label{Eq:CoxRL_N_11}
\end{eqnarray}
with $m=1,2,3,\ldots,N-1$ and $n=1,2,3,\ldots,N-1$.

Here one sees the emergence of hypergeometric distribution (Eq.~\ref{Eq:HypergeometricDistribution}) and of the couple density $P^{(rl, N)}_{1, 1}$.
These equations (Eqs.~\ref{Eq:CoxRL_N} and~\ref{Eq:CoxRL_N_11}) reduce to Eqs.~\ref{Eq:CoxRandomLink} and~\ref{Eq:CoupleDensityRL} as  $N \gg 1$.
 
\begin{figure}[htb]
\begin{center}
\includegraphics[angle=-90, scale=.5]{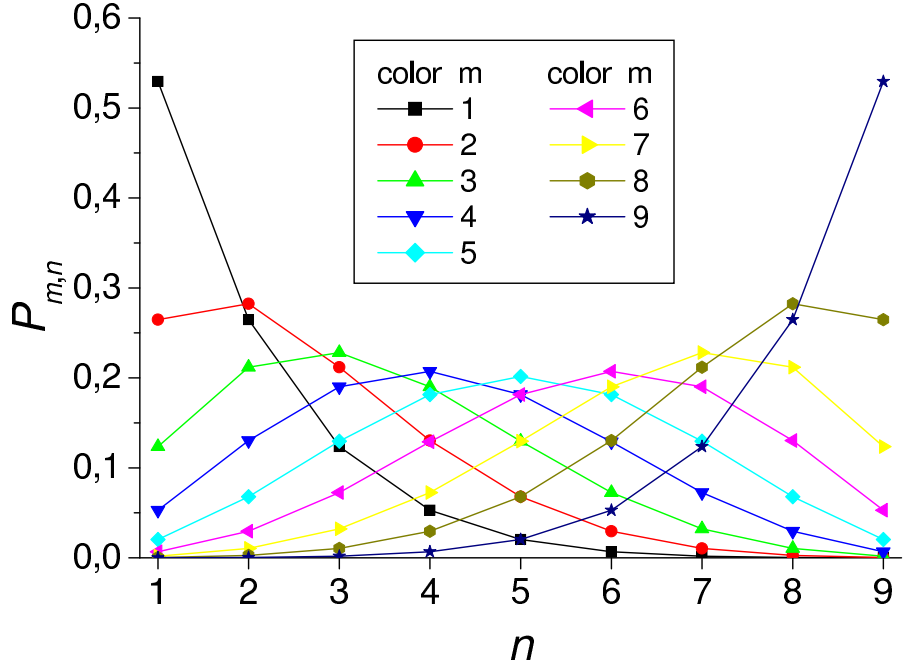}
\caption{Neighborhood probabilities in a 10-point RLM.
The distributions are discrete and the lines are only a guide to the eyes.} 
\label{Fig:RL}
\end{center}
\end{figure}

Fig.~\ref{Fig:RL} shows $P^{(rl, N)}_{m, n}$ as a function of $n$ in a 10-point RLM.
Notice that each curve reaches its maximum at the reflexive case $m=n$ and that they are symmetric with respect to $N/2$.

\section{Random Map Model}
\label{sec:rmm} 

Breaking the distance symmetry constraint $D_{i, j} = D_{j, i}$ in the RLM leads to the RMM.
The RMM is the mean field approximation to several problems and analytical results may be obtained.
Also, Cox probabilities can be obtained for the RMM. 

In the case which the constraint $D_{i, i} = 0, \forall i$ is preserved, if an arbitrary point $I$ is chosen, its $m$th neighbor $J$ is automatically set, but the $n$th neighbor of $J$ is equally probable to be anyone of the other $N-1$ points, since the distances are totally independent.
Thus, the probability $P^{(rm)}_{m,n}$ that the point $I$ is the $n$th neighbor of its $m$th neighbor is simply:
\begin{equation}
P^{(rm)}_{m,n}  = \frac{1}{N-1} \; ,
\end{equation}
where $m = 1, 2, \ldots, N-1$ and $n = 1, 2, \ldots, N-1$.

On the other hand, in the case which $D_{i, i} \ne 0$ is allowed, the probability $P^{(rm)}_{m,n}$ is twice as large for reflexive neighbors than for non-reflexive ones, because now one must consider that every point is always its own $m$th nearest neighbor, for some $m$.
\begin{equation}
P^{(rm)}_{m,n}  = \frac{1 + \delta_{m, n}}{N+1} \; ,
\end{equation}
where $\delta_{m, n}$ is the Kronecker delta, $m = 1, 2, \ldots, N$ and $n = 1, 2, \ldots, N$.

Notice that in the thermodynamic limit $N \gg 1$, these cases are still distinguable due to the presence  of the factor 2 for the reflexive neighbors. 

\section{Conclusion}
\label{sec:conc}

Using only Poisson distribution, Cox probabilities have been obtained through a simple derivation and they have been identified with the multinomial distribution.
Writing the dimensionality parameter $p_d$ in terms of the normalized incomplete beta function allowed us to obtain the high dimensional approximation for the neighborhood probabilities in Poissonic processes (RPP, for instance) and a characteristic dimensionality $d_0=8$ has arisen naturally.

Using the same line of reasoning, the neighborhood probabilities have been obtained for RLM finite size systems. 
In this case the probabilities have been identified with the hypergeometric distribution.
Also, simple expressions have been obtained for the RMM.

Up to now, we are devoting efforts to try to obtain the neighborhood probabilities for finite-size and low-dimensionality systems. 

\ack

The authors thank the discussions with R. S. Gonz\'alez and U. P. C. Neves. 
ASM acknowledges the support from the Brazilian agencies CNPq (305527/2004-5) and FAPESP (2005/02408-0). 

%
\setcounter{section}{1}
\section*{Appendix: Some Discrete Statistical Distributions}
\label{sec:appendix}

In the following we briefly review some discrete distributions used here. 

\subsection{Infinite Population}

Let us first consider infinite populations.

\subsubsection{Multinomial Distribution}

Consider an infinite population whose objects can be classified according to $m$ distinct types, which occur with probabilities $p_1, p_2, \ldots, p_m$, such that $\sum_{i=1}^m p_i = 1$. 
The probability that a uniform random sample has $n_1$ objects of type $1$, $n_2$ objects of type $2$ and so on is given by the \emph{multinomial distribution}:
\begin{equation}
\mbox{mult}(n_1, n_2, \ldots, n_m; p_1, p_2, \ldots, p_m) = \frac{n!\; p_1^{n_1} p_2^{n_2} \cdots p_m^{n_m}}{n_1! n_2! \cdots  n_m!}  \; ,
\label{Eq:MultinomialDistribution}
\end{equation}
where $n = n_1 + n_2 + \cdots + n_m$ is the sample size.

\subsubsection{Binomial Distribution}

If the infinite population has only two distinct types of objects (usually referred as success and failure), the multinomial distribution becomes the \emph{binomial distribution}:
\begin{equation}
\mbox{bin}(n_1, n_2; p_1, p_2)  =  \frac{n!p_1^{n_1} p_2^{n_2}}{n_1! n_2!} = \left(\begin{array}{c} n \\ n_1 \end{array}\right) \; p_1^{n_1} (1 - p_1)^{n - n_1} \; ,
\label{Eq:BinomialDistribution}
\end{equation}
where $n = n_1 + n_2$ and $p_2 = 1 - p_1$. 

\subsubsection{Poisson Distribution}

If $n \rightarrow \infty$ and $p_1 \rightarrow 0$ such that the average $\mu = n p_1$ remains finite, the binomial distribution becomes the \emph{Poisson distribution}:
\begin{equation}
\mbox{pois}(n_1) =  \frac{\mu^{n_1} e^{-\mu}}{n_1!} \; ,
\label{Eq:PoissonDistribution}
\end{equation}
where the only distribution parameter is the average $\mu = \langle n_1 \rangle$.

\subsection{Finite Population}

Consider a finite population with $N$ objects, such as $N_1$ objects are of type $1$ and the reminding $N_2=N-N_1$ objects are of type $2$. 
The probability that a uniform random sample, drawn without reposition, has $n_1$ objects of type $1$, $n_2$ objects of type $2$ and so on is given by the \emph{hypergeometric distribution}:
\begin{equation}
\mbox{hypg}(N_1, N_2; n_1, n_2) = \frac{\left( \begin{array}{c} N_1 \\ n_1 \end{array} \right) \left( \begin{array}{c} N_2 \\ n_2 \end{array} \right)}{\left( \begin{array}{c} N \\ n \end{array} \right)} \; ,
\label{Eq:HypergeometricDistribution}
\end{equation}
where $n = n_1 + n_2$ is the sample size.
In the limit $N_1 \rightarrow \infty$ and $N_2 \rightarrow \infty$, with the probabilities $p_1 = N_1/N$ and $p_2 = N_2/N$ kept fixed, the hypergeometric distribution becomes the binomial one.

\end{document}